\documentclass[journal=jpca,manuscript=article]{achemso}

\usepackage{ragged2e}
\usepackage{graphicx}
\usepackage{color}
\usepackage{hyperref}
\usepackage{float}
\usepackage{array, multirow}
\usepackage{amsmath, amssymb}
\usepackage{newtxtext}              
\usepackage{booktabs}
\usepackage{xcolor} 
\usepackage{mathtools}
\usepackage{soul}
\usepackage{url}
\usepackage{qcircuit}
\usepackage{ulem}
\usepackage[T1]{fontenc}
\usepackage{subcaption}

\hypersetup{colorlinks, 
    linkcolor={blue!75!black!80!yellow},
    citecolor={blue!75!black!80!yellow}, 
    urlcolor={blue!75!black!80!yellow}
    }

\title{Quantum Configuration Interaction with a Partial Walsh Series}

\author{Koray Aydo\u{g}an}
\affiliation{School of Physics and Astronomy, University of Minnesota, Minneapolis, MN 55455 USA}
\author{Anna R. Spak}
\affiliation{Department of Mathematics, University of Rochester, Rochester, NY 14627 USA}
\author{Kade Head-Marsden}
\affiliation{Department of Chemistry, University of Minnesota, Minneapolis, MN 55455 USA}
\alsoaffiliation{School of Physics and Astronomy, University of Minnesota, Minneapolis, MN 55455 USA}
\author{Anthony W. Schlimgen}
\affiliation{Department of Chemistry, University of Minnesota, Minneapolis, MN 55455 USA}

\email{aws@umn.edu}

\begin{document}

\maketitle

\begin{abstract}
Configuration interaction (CI) is a foundational approach in modern quantum chemistry for strongly correlated wavefunctions; however, development of post-CI quantum algorithms has been hampered by challenges encoding the initially correlated wavefunction. Here we describe an exact and surjective encoding of CI wavefunctions into a probabilistic quantum circuit using a subspace superposition, then applying diagonal Walsh operators to prepare the wavefunction. The transformation can be used as a quantum \textit{Ansatz} for full CI and selected CI wavefunctions, resulting in exact and near-exact solutions for electronic ground states. We demonstrate results for several molecules using quantum simulators and hardware. By sampling the Walsh basis, the encoding bypasses classical matrix diagonalizations, which is advantageous for large-scale applications. This simplified encoding of CI wavefunctions enhances the possibilities for quantum algorithms for post-CI methods such as multireference coupled-cluster and perturbation theories.
\end{abstract}

\textit{Introduction--} 
Multireference (MR) quantum chemistry provides detailed and quantitative physical insight into the behavior and properties of many molecular quantum systems.~\cite{Bartlett:2007,Angeli:2001} The complexity of these systems grows unfavorably with system size which limits the complete description of electron correlation for many interesting species. In the context of quantum chemistry and electronic structure theory, strongly correlated wavefunctions correspond to states that are superpositions of Slater determinants or configuration state functions (SDs);~\cite{Szabo:2012} however, quantitative description of strongly correlated, or MR, electronic wavefunctions remains a significant challenge for classical algorithms.~\cite{Szalay:2012,Wardzala:2026aa} Frequently a qualitatively correct MR state is computed, but the energy and properties must be corrected by the dynamical correlation energy for quantitative accuracy. \textcolor{black}{It is common to describe electron correlation in terms of various classes or types, but here we consider recovering the  dynamical correlation energy beyond an MR initial state.~\cite{Angeli:2001, Szalay:2012, Gagliardi:2017}} The cost of the dynamical correlation correction scales unfavorably with the degree of correlation in the intial wavefunction, which limits the applicability of MR quantum chemistry in classical simulations. \textcolor{black}{The design of classical and quantum algorithms for computing MR dynamical correlation is essential for the progress of quantitative quantum chemistry.}

This computational challenge inspires continued improvements in simulating quantitatively correct electronic states using quantum and classical algorithms,~\cite{Abrams:1997, Abrams:1999, McArdle:2019, Aspuru-Guzik:2005, Lanyon:2010, McArdle:2020, OMalley:2016, Peruzzo:2014, McClean:2016, Holmes:2016, White:1999, Foulkes:2001, Angeli:2001, Bartlett:2007, Chan:2024} with significant interest in strongly correlated wavefunctions~\cite{Szalay:2012, LeBlanc:2015, Gagliardi:2017, Paschen:2021}. Quantum algorithms offer asymptotic advantages for MR wavefunctions, but may require large-scale error-correcting or fault-tolerant quantum computers to be competitive platforms for practical quantum simulation~\cite{Katabarwa:2024}. Several standard quantum chemistry methods have been implemented as quantum algorithms, including Hartree-Fock, imaginary-time evolution, and coupled-cluster (CC), among others~\cite{Motta:2020aa, Kamakari:2022, Hejazi:2024, Baek:2023}. On the other hand explicilty MR quantum algorithms such as multireference CI (MRCI) and multireference CC (MRCC) have received less attention~\cite{Stair:2020, Greene-Diniz:2021, Wu:2025}. This is partially due to the complexity of the initial states which are, by construction, a superposition of SDs. Efficient quantum encodings of MR initial states are vital due to the immense improvement of classical approximate CI techniques, which greatly expands the tractability of CI approaches for electronic structure~\cite{Parrish:2019, Shirai:2025, Verma:2026}. These approximate techniques yield high-quality, but approximate, MR wavefunctions which could be used as inital states for quantitive correction via quantum algorithms.

One of the most commonly employed quantum algorithms is the variational quantum eigensolver (VQE), which iteratively minimizes the ground-state energy of a paramerized quantum circuit, known as an \textit{Ansatz}. Significiant progress has been made developing resource-efficient and physically-motivated VQE \textit{Ans\"atze} for molecular Hamiltonians~\cite{Kandala:2017, McClean:2016, Zhang:2021, Zhang:2022, Grimsley:2019, Yordanov:2021, Tang:2021, Anastasiou:2024, Ramoa:2025}; however, challenges in standard \textit{Ans\"atze} include the likelihood of overparameterization and barren plateaus, which are flat areas of the potential surface that stymie optimization progress~\cite{McClean:2018aa, Larocca:2025, Qi:2026}. VQE typically employs the Hartree-Fock reference as the initial state, which simplifies state-preparation, but can lead to deep circuits for strongly-correlated wavefunctions. Several methods have been introduced to remedy this problem, including active-space approaches,~\cite{Wang:2008, DCunha:2024} sums of Slater determinants,~\cite{Babbush:2015, Tubman:2018} and preparation of matrix-product states,~\cite{Berry:2025} along with quantum read-only memory techniques.~\cite{Berry:2019, Low:2024, Fomichev:2024} In spite of this, the state-preparation problem, overparameterizations, and barren plateaus in VQE \textit{Ans\"atze} continue to prevent effective implementation of explictily MR quantum algorithms.

Here we describe a mapping for CI wavefunctions that simplifies both the quantum state preparation and circuit parameterizations for MR wavefunctions. By employing Walsh functions, we derive a surjective and exact mapping between the CI coefficients and rotation angles implemented in a diagonal circuit, such that the circuit is not overparameterized. The initial wavefunction can be easily augmented to include more variational parameters, which can be optimized with VQE or other quantum algorithms, resulting in a family of quantum MRCI algorithms that are analogous to their classical counterparts. In addition, because the \textit{Ansatz} is diagonal, all of the circuit layers commute, avoiding barren plateaus induced by approximate 2-designs~\cite{McClean:2016, Cerezo:2021, Ragone:2024, Qi:2026}. This algorithm is probabilistic, but requires only one ancilla qubit regardless of system size. We demonstrate this strategy by performing quantum variants of full CI (FCI), MRCI singles and doubles (MRCISD), and selected CI (SCI). We compute the ground-state energy for a range of molecules with IBMQ's Torino processor and simulator. \textcolor{black}{Our results demonstrate a bona fide MR quantum algorithm and provides a path for development of more explicitly MR quantum algorithms. These approaches can provide corrections to the dynamical correlation energy, which is essential for quantitative accuracy in contemporary quantum chemistry.}

\textit{Theory and Methods--} A molecular electronic Hamiltonian under the Born-Oppenheimer approximation can be expressed as~\cite{Szabo:2012}, 
\begin{align}
    \hat{H} = \sum_{mn} K^m_n \hat{a}^{\dagger}_m \hat{a}_n + \sum_{mnpq} V^{mn}_{pq} \hat{a}^{\dagger}_m \hat{a}^{\dagger}_n \hat{a}_q \hat{a}_p,
\end{align}
where $\hat{a}$ and $\hat{a}^{\dagger}$ are fermionic annihilation and creation operators, and $K^m_n$ and $V^{mn}_{pq}$ are one- and two-body integrals, respectively. For this Hamiltonian, a general $N$-electron wavefunction, $|\Psi \rangle$, can be written as a linear combination of SDs, 
\begin{equation}
    |\Psi \rangle = \sum_{k=1}^D c_k |\psi_k\rangle,
    \label{eq:wf}
\end{equation}
where $D$ is the number of SDs in the expansion, and $c_k$ and $|\psi_k\rangle$ are the $k^\textrm{th}$ coefficient and $N$-electron SD, respectively. Generally called a CI wavefunction, the choice of the set of SDs results in approximate eigenstates for the Hamiltonian in $r$ spin-orbitals or qubits. For example, full CI (FCI) uses the expansion that includes all $N$-electron basis functions, and can represent the exact eigenstates of the Hamiltonian~\cite{Szabo:2012}. Such an approach scales combinatorially in $r$ and $N$; however, \textcolor{black}{recent advances in finding accurate and approximation CI expansions based on SCI have significantly expanded the capacity of CI techniques beyond standard FCI algorithms}.~\cite{Booth:2013, Holmes:2016}

Enconding these wavefunctions in a quantum circuit is non-trivial for at least two reasons. First, while Fock space is constructed to enforce the symmetries of $N$-electron wavefunctions, the full qubit Hilbert space does not. Second, MR wavefunctions tend to have amplitudes that span the full qubit space, resulting in deep circuits for encoding. These two considerations motivate the two steps of our algorithm: superposition preparation, and \textit{Ansatz} application. Because the $r$-qubit Hilbert space contains many more Slater determanints than the $N$-electron space, we prepare the uniform superposition only over \textit{selected} determinants, which could correspond to FCI or some other expansion. Finally, we apply a diagonal operator of CI coefficients to the uniform superposition to prepare the CI wavefunction. This diagonal operator is not unitary in general, so we use an ancillary qubit and construct a dilated diagonal unitary through a simple transformation.  

A uniform superposition on a quantum circuit can be prepared in a variety of ways depending on the structure of the superposition. For example, the superposition over all bitstrings requires only Hadamard gates on each qubit, but this will overparameterize an $N$-electron wavefunction. Alternatively, we can prepare superpositions of Dicke states, which are number-conserving in the qubit basis.~\cite{Dicke:1954} Dicke states can be prepared without ancilla qubits with CNOT-scaling $\mathcal{O}(rN)$.~\cite{Nielsen:2010, Bartschi:2019} We can also consider arbitrary CI expansions which do not have any particular structure, and can be prepared via quantum walks with CNOT scaling $\mathcal{O}(rD)$.~\cite{Gonzales:2025} The superposition step allows us to restrict the parameter space of the $r$-qubit wavefunction and encode physical symmetries of the CI wavefunction by appropriate choices of superposed basis functions.

In general, we denote the subspace preparation as $U_S$,
\begin{equation}
    |S\rangle = U_S|0\rangle^{\otimes r},
\end{equation}
where $|S\rangle$ is the superposed subspace, and $|0\rangle^{\otimes r}$ indicates the $r$-fold tensor product of the ground state. Having prepared the superposition, we then encode the initial CI wavefunction. Following previous work for CI coefficients $c_k$ in Eq.~\ref{eq:wf} , we define a diagonal unitary matrix,
\begin{align}
    U &= \begin{pmatrix}
    \Sigma_+ & 0 \\
    0 & \Sigma_-
    \end{pmatrix} ; \quad \Sigma_{\pm k} = c_k \pm ic_k\sqrt{\frac{1 - \Vert c_k \Vert^2}{\Vert c_k \Vert^2}},
\end{align}
where $\Vert c_k \Vert$ is the norm of $c_k$,~\cite{Schlimgen:2022aa} $\Sigma_{\pm k}$ is a unit magnitude complex scalar for all $k$, and thus $\Sigma_\pm$ are unitary diagonal matrices. We emphasize that ${2\Sigma = \Sigma_+ + \Sigma_-}$, which is a diagonal matrix with $c_k$ on the diagonal. After applying $U_S$, we apply the Hadamard gate to the ancilla, followed by $U$, which acts over all $r+1$ qubits. Finally, we apply a Hadamard to the ancilla again. When the ancilla is measured in state $|0\rangle$ we prepare,
\begin{equation}
    \frac{1}{2}(\Sigma_+ + \Sigma_-)|S\rangle = \frac{1}{2}(2\Sigma)|S\rangle = |\Psi\rangle,
\end{equation}
and when the ancilla is in state $|1\rangle$, the negative combination of $\Sigma_{\pm}$ is prepared. Each of the wavefunctions conditioned on the ancilla outcome have unit norm, implying each outcome will occur exactly half of the time.

\begin{figure}
    \centering
    \includegraphics[width=0.48\textwidth]{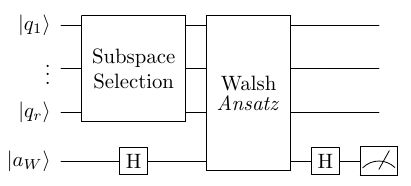}
    \caption{Circuit diagram for the preparation of $|\Psi\rangle$, where H is the Hadamard gate and $|a_W\rangle$ is the ancilla qubit for the diagonal Walsh \textit{Ansatz}. The subspace selection step prepares a selected uniform superposition.}
    \label{fig:walsh_circuit_blocks}
\end{figure}

The dilated diagonal unitary $U$ can be implemented with Walsh operators as,
\begin{equation}
    U = \prod_{j ~\textrm{odd}} e^{i a_j \hat{w}_j},
    \label{eq:walsh_u}
\end{equation}
where $a_j$'s are the Walsh coefficients, $\hat{w}_j$ are the $j^\textrm{th}$ order Walsh operators, and the product is over the entire system and ancilla space, $2^{r+1}$.~\cite{Golubov:1991} Because $\Sigma_+$ and $\Sigma_-$ are complex conjugates, only half of the Walsh operators are needed. This operator can be implemented on a quantum circuit with CNOT and $R_Z$ gates using the binary expansion of $j$, most efficiently ordered by the Gray code.~\cite{Nielsen:2010, Welch:2014} A schematic of the entire circuit is shown in Figure~\ref{fig:walsh_circuit_blocks}.
We express the Walsh coefficients $a_j$ directly in terms of the wavefunction coefficients $c_k$,
\begin{equation}
\begin{aligned}
    a_j =  2^{-r}~\sum_{k=1}^{2^{r}} c_k (-1)^{\langle k.j\rangle},
    \label{eq:walsh_coeff}
\end{aligned}
\end{equation}
where $\langle k. j\rangle$ is the binary inner product of the integer representations of the $k^{th}$ SD and Walsh order $j$.  As noted above, $N$-electron wavefunctions do not span the complete $2^r$ Hilbert space, so the sum in Eq.~\ref{eq:walsh_coeff} must be restricted to only those $D$ coefficients which are present in the chosen wavefunction expansion in Eq.~\ref{eq:wf}. This also requires that the subspace superposition is prepared only over $D$ chosen bitstrings.

While mapping symmetry-preserving SDs to symmetry-preserving bitstrings can be achieved either using Dicke state preparation or quantum walk methods, it is not generally necessary to match the occupation patterns between SDs and the bitstrings in the computational basis. In fact, there may be several selections of $D$ Walsh functions that can provide a full-rank Walsh-Fourier transform (WFT) of $D$ SDs, though an arbitrary selection Walsh functions may not result in a full-rank WFT. To overcome this, we can oversample the Walsh basis by selecting $\mathcal{O}(D\textrm{log}D)$ functions randomly from the uniform distribution, which generates a full-rank WFT with probability scaling inversely with system size, $\mathcal{O}(1 - D^{-1})$.~\cite{Tropp:2011} The randomly selected Walsh functions $\hat{w}_j$ are then used in Eq.~\ref{eq:walsh_u} and coefficients $a_j$ are found with Eq.~\ref{eq:walsh_coeff}. Oversampling the Walsh basis results in an overparameterized circuit, because there are $\mathcal{O}(D\textrm{log}D)$ circuit parameters and only $D$ wavefunction parameters, but this limits the classical preprocessing cost to matrix-vector multiplication. Importantly, while this results in a trade-off between classical cost and number of parameters, the overparameterization scales at worst \textit{linearly} with system size $r$, because $D \propto 2^r$ in the worst case. In benchmark NISQ examples, we can also find $D$ orthogonal Walsh functions from an oversampled set using the QR decomposition, which scales as $\mathcal{O}(D^3)$ and preserves the surjectivity of the \textit{Ansatz}.~\cite{Virtanen:2020} We use both the exactly parameterized and oversampled approaches in the examples that follow.

\textit{Results--} We find the ground states of various molecules using VQE with either FCI or SCI, and benchmark the simulations with classical FCI and coupled-cluster single-doubles with perturbative triples (CCSD(T)) using PySCF.~\cite{Szabo:2012, Bartlett:2007, Sun:2018} We obtain the electronic Hamiltonian from OpenFermion, transformed to the qubit basis using the Jordan-Wigner transformation,~\cite{Whitfield:2011, McClean:2020} and use Qiskit to construct the quantum circuits for the statevector simulator, quantum device simulator, and IBMQ's Torino quantum hardware, with details specified in the Supplementary Information.~\cite{Sun:2018, Javadi-Abhari:2024, Abughanem:2025} To estimate the energy we use the Hadamard test to measure the expectation values of Pauli strings, requiring another ancilla qubit.~\cite{Aharonov:2009} For H$_2$, we converge the energy to chemical accuracy, $1.6\times 10^{-3}$ Hartree, and H$_6$ we converge the energy to $10^{-5}$ Hartree, and use the COBYLA and BFGS optimizers implemented in SciPy respectively.~\cite{Virtanen:2020} For the device and noisy simulator we sample with 1024 measurements. We find a full rank WFT by oversampling the Walsh basis and use the QR decomposition to find a $D$ by $D$ WFT. 

\begin{figure}[h!]
    \centering
    \begin{subfigure}[b]{0.48\textwidth}
        \centering
        \includegraphics[width=\textwidth]{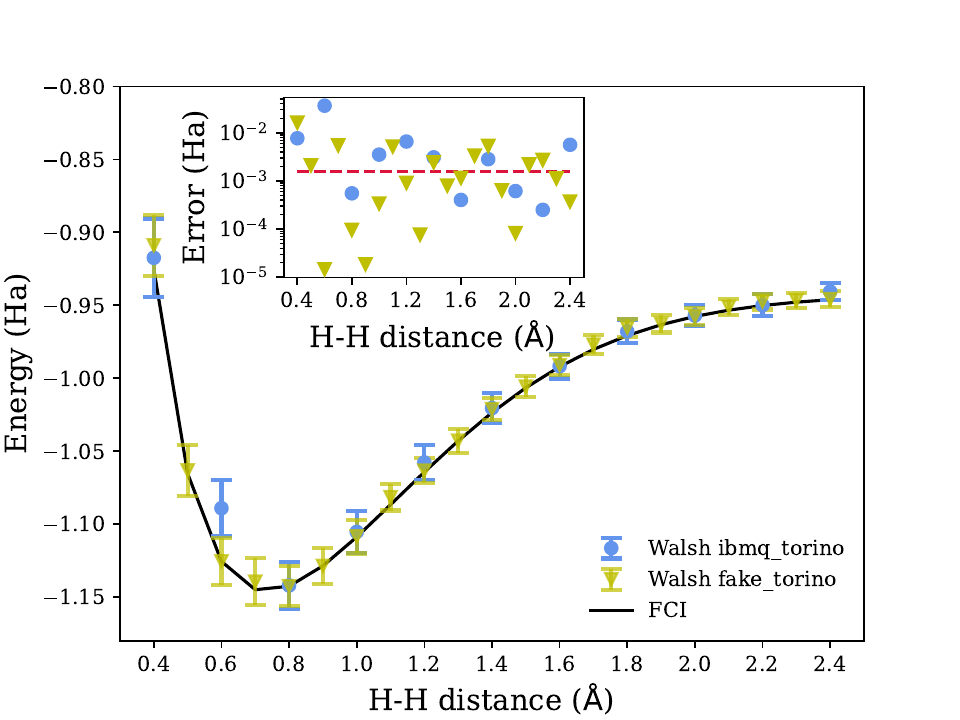}
        \put(-210, 162){\textbf{(a)}} 
        \label{fig:H2_torino}
    \end{subfigure}
    \hspace{1mm}
    \begin{subfigure}[b]{0.48\textwidth}
        \centering
        \includegraphics[width=\textwidth]{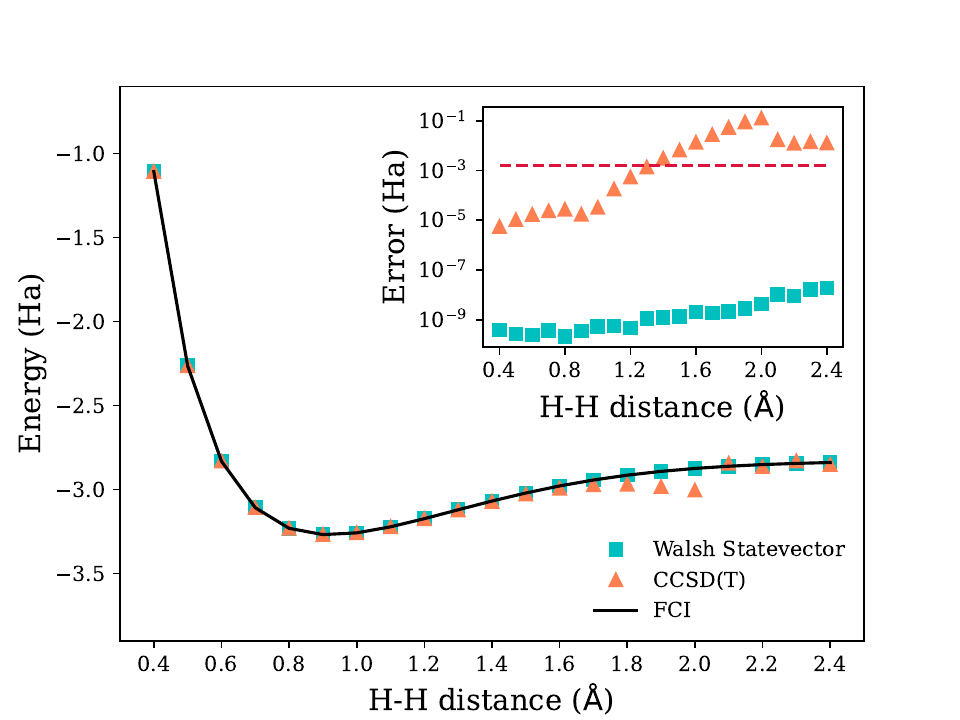}
        \put(-210, 162){\textbf{(b)}} 
        \label{fig:H6_methods}
    \end{subfigure}
    \caption{Dissociation of (\textbf{a}) H$_2$ using the subspace selected Walsh approach on IBMQ's Torino processor (blue circles), Torino noisy simulator (green triangles) each with 1024 shots, and the FCI solution (line) and (\textbf{b}) linear H$_6$ with a statevector simulator (teal squares), CCSD(T) (orange triangles), and FCI (black line). The red dashed line denotes 1.6 mHa error, which is chemical accuracy. \textcolor{black}{Error bars represent the mean statistical uncertainty across all VQE optimization iterations on IBMQ's Torino processor and Torino noisy simulator.}}
    \label{fig:H2_and_H6}
\end{figure}

In Figure~\ref{fig:H2_and_H6} we show the the dissocation curves of the quantum FCI solutions for H$_2$ and H$_6$ in a minimal STO-6G basis.~\cite{Hehre:1969} 
For H$_2$ we use quantum walk to prepare the superposition, conserving spin- and particle-number symmetries, which results in two SDs for the FCI wavefunction, $|1100\rangle$ and $|0011\rangle$~\cite{Szabo:2012}. Figure~\ref{fig:H2_and_H6}~(a) shows the dissociation of H$_2$ computed with FCI (black line), the Torino noisy simulator (blue circles), and the Torino quantum device (green triangles). The device and the simulator data are in excellent agreement with the FCI solution, almost always surpassing chemical accuracy. The dissociation curve of H$_6$ is shown in Fig.~\ref{fig:H2_and_H6}~(b). We compare the results of the statevector simulator (teal squares), CCSD(T) (orange triangles), and FCI (black line). Here we include all spin preserving $N$-electron SDs, resulting in 400 parameters or SDs, and prepare the superposition with quantum walks. The ideal statevector simulator results highlight the accuracy of the present algorithm. 

\begin{figure}[h!]
    \centering
    \includegraphics[width=0.48\textwidth]{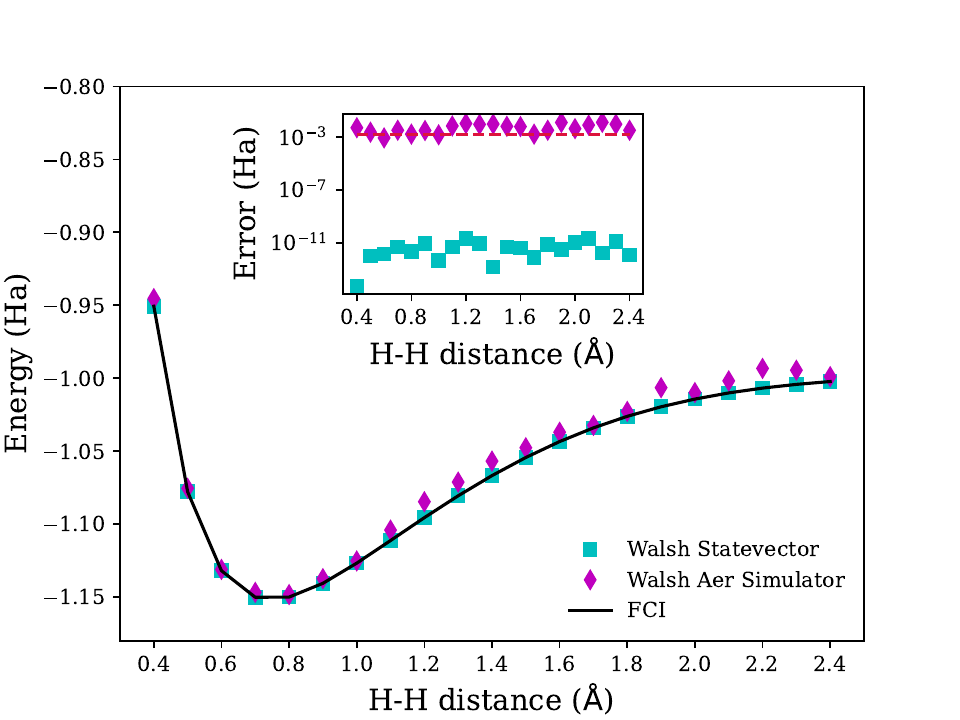}
    \caption{Dissociation of H$_2$ in 6-31G basis with a statevector simulator (teal squares), noiseless Aer simulator (purple diamonds), using an oversampled Walsh basis. The red dashed line denotes 1.6 mHa error, which is chemical accuracy.}
    \label{fig:h2-631g}
\end{figure}

For large-scale applications and molecules in non-minimal basis sets, it is advantageous to avoid the classical QR decomposition. As described above we randomly oversample the Walsh basis with $D\textrm{log}D$ functions. This tradeoff between the the number of CNOT gates and the parameters of the \textit{Ansatz} and classical resource efficiency provides a practical approach for systems which are not possible with a QR decomposition. As an example we use H$_2$ in a 6-31G basis set resulting in 44 circuit parameters for 16 SDs, shown in Figure~\ref{fig:h2-631g}.~\cite{Ditchfield:1971aa} We show results from the statevector simulator (teal squares) and noiseless Aer sampler (purple diamonds) using 10 qubits and $2^{18}$ shots.

\begin{figure}[h!]
    \centering
    \includegraphics[width=0.48\textwidth]{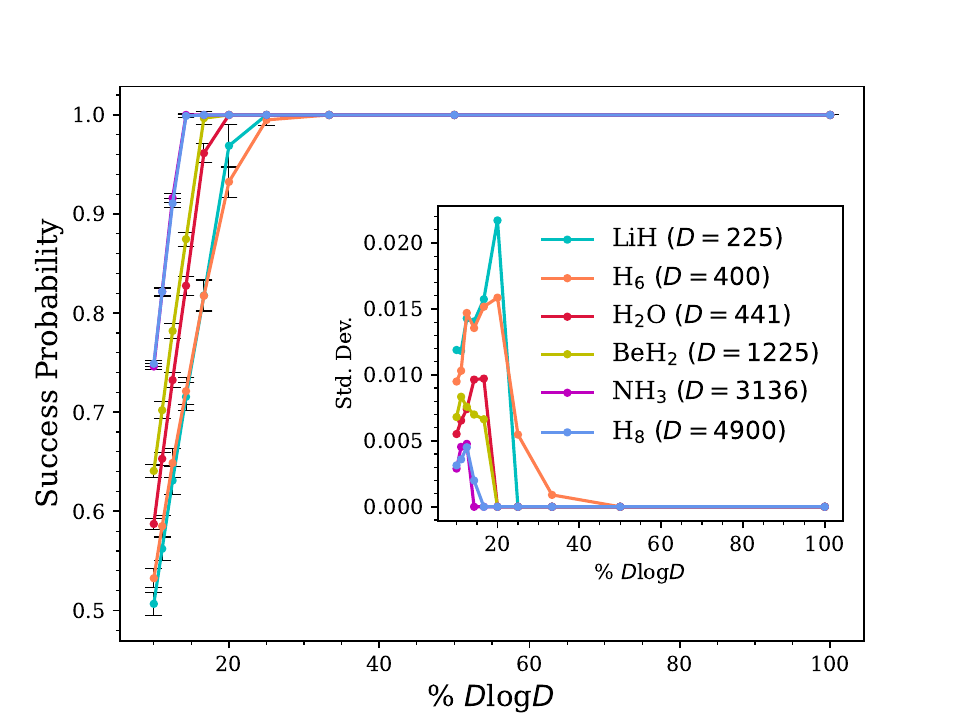}
    \caption{The median success probability of obtaining a full-rank WFT for several molecules, using fractions of $D\textrm{log}D$ randomly selected Walsh functions. Error bars represent the standard deviation of 50 trials.} 
    \label{fig:random_fraction}
\end{figure}

We also investigate whether all $D\textrm{log}D$ oversampled Walsh functions are necessary for a full rank transformation. Figure~\ref{fig:random_fraction} shows the likelihood of finding a full rank transformation by random selection as a function of the fraction of $D\textrm{log}D$ for several molecules. To generate the data, we perform 50 random selections of Walsh functions from the uniform distribution, find the rank of the transform, and compute the success probability as the computed rank divided by the full rank. We find that for all molecules we can find a full rank transformation by random selection at about 50\% of $D\textrm{log}D$. As the system size increases, a smaller fraction of $D\textrm{log}D$ Walsh functions is required to achieve a full rank transformation, which agrees with the previously reported $\mathcal{O}(1-D^{-1})$ results.~\cite{Tropp:2011} As expected, as system size increases, the required oversampling rate becomes smaller. 

We investigate the gate scaling of our quantum FCI approach by considering a series of molecules with $N$ electrons, $D$ SDs, and $r$ qubits. The subspace selection is achieved by either Dicke or quantum walk state preparation, whose CNOT gate scalings are $\mathcal{O}(rN)$ and $\mathcal{O}(rD)$, respectively. In Table~\ref{tab:molecules_state_prep_walsh_gates} we show the number of CNOT gates for both the state preparations and the Walsh components with H$_6$ and H$_8$ chains, along with LiH, BeH$_2$ and NH$_3$ molecules. 
\begin{table*}[htb!]
    \centering
    \resizebox{\textwidth}{!}{
    \begin{tabular}{@{\extracolsep{4pt}}l c c c c c} 
        \hline \hline
         \multicolumn{1}{c}{Symmetry} &  \text{H$_6$ (12 qubits)}   & \text{H$_8$ (16 qubits)} & \text{LiH (12 qubits)} & \text{BeH$_2$ (14 qubits)}  & \text{NH$_3$ (16 qubits)} \\
         \cline{1-1}\cline{2-2}\cline{3-3}\cline{4-4}\cline{5-5}\cline{6-6}

         Spin \& number & \text{52 + 726 = 778} &  \text{391 + 8756 = 9147}  &  \text{96 + 520 = 616}  &  \text{108 + 2328 = 2436}  &  \text{218 + 7530 = 7748}\\
         
         Number & \text{382 + 1622 = 2004}  &  \text{723 + 22652 = 23375}  &  \text{265 + 1154 = 1419}  &  \text{476 + 6292 = 6768}   &  \text{840 + 18334 = 19174}\\
        
         No symmetry & \text{0 + 4118 = 4118}  &  \text{0 + 65566 = 65566} &  \text{0 + 4118 = 4118}  &  \text{0 + 16410 = 16410}  &   \text{0 + 65566 = 65566} \\
        \hline \hline
    \end{tabular}
    }
    \caption{Number of CNOT gates for several molecules, where each row imposes different subspace symmetries. The first and the second numbers are the CNOT gate count for subspace selection and the Walsh \textit{Ansatz}, respectively. Spin and particle-number conservation is enforced by the quantum walk algorithm; particle number conservation alone employs the Dicke state preparation algorithm. We show the number of system qubits, which is supplemented by two ancilla qubits.}
    \label{tab:molecules_state_prep_walsh_gates}
\end{table*}
In general, our resource requirements are similar to those of recent high-quality ADAPT-VQE circuits.~\cite{Anastasiou:2024, Ramoa:2025} Here, we find the exact FCI solution under three different cases: spin and particle-number conserving, particle-number conserving, and no symmetry. 

The Walsh \textit{Ansatz} has favorable scaling in both the two-qubit CNOT gates, and the number of variational parameters required for the one-qubit rotation gates with respect to the size of the parameter space. The number of CNOT gates for the Walsh \textit{Ansatz} scales linearly with the number of determinants chosen, $\mathcal{O}(D)$. Furthermore, the Walsh \textit{Ansatz} has the same number of variational parameters for $R_Z$ gates as the number of SDs in the selected expansion. This highlights that the \textit{Ansatz} introduces no overparameterization of the chosen subspace, which helps in limiting spurious local minima in the optimization. 

If we choose to represent the FCI wavefunction then the circuit will require exponentially scaling parameters and CNOT gates. Instead, we may select physically relevant CI coeffients using a variety of procedures. For example, starting from a classically computed CI expansion we can include excitations connected through a given order of perturbation theory or through a heuristic. The former procedure is standard in coupled-cluster and multireference perturbation theories,~\cite{Bartlett:2007, Angeli:2001} and the latter is standard in classical SCI approaches.~\cite{Tubman:2016, Schriber:2016, Holmes:2016}

We adapt the SCI approach to our quantum algorithm, using H$_2$O in an STO-6G basis with a classically computed complete active space configuration interation (CASCI) wavefunction composed of 8 electrons in 10 spin orbitals.~\cite{Szabo:2012} From this wavefunction, we retain CI coefficients above a threshold $\epsilon$, then add SDs corresponding to single and double excitations from this truncated reference, which constitutes the variational wavefunction for the Walsh VQE. \textcolor{black}{We essentially generate an MRCISD wavefunction from a classicaly computed truncated CASSCF reference.} In this case we prepare the number- and spin-preserving subspace with the quantum walk algorithm. Table~\ref{tab:H2O_threshold_energies} shows the results for H$_2$O at the equilibrium geometry, $r_{\textrm{OH}}=0.96$~{Å} and $\theta_{\textrm{OHO}}=104.3^\circ$, with two different $\epsilon$ thresholds and quantum FCI using a statevector simulator. We list the number of SDs, the number of CNOT gates, the resulting energy, and the fidelity compared to the exact solution. As expected, we see systematic improvement in the fidelity as the SCI space is increased.
\begin{table}[H]
    \centering
    \begin{tabular}{@{\extracolsep{4pt}}l c c c} 
        \hline \hline
         \multicolumn{1}{c}{H$_2$O (14 qubits)} &  \text{SCI ($\epsilon = 10^{-3}$)}  & \text{SCI ($\epsilon = 10^{-7}$)}  & \text{FCI}\\
         \cline{1-1}\cline{2-2}\cline{3-3}\cline{4-4}
         
        \# of SDs & \text{227} & \text{267} &  \text{441} \\
        \# of CNOTs (SS) & \text{579}  & \text{378} & \text{60}  \\
        \# of CNOTs (W) & \text{722}  & \text{784} & \text{1352}  \\
        Energy Error (Ha) & \text{0.031} &  \text{0.020}  &  \text{0.000} \\
        Fidelity & \text{0.991}  & \text{0.994}  &  \text{1.000} \\
        \hline \hline
    \end{tabular}
    \caption{Number of Slater determinants with two different thresholds for the SCI and quantum FCI, number of CNOT gates, energy error, and fidelity with FCI for H$_2$O at its equilibrium geometry. (SS) and (W) are quantum walk subspace selection and Walsh \textit{Ansatz}, respectively. The quantum FCI result matches the classical result to machine precision. We show the number of system qubits, which needs to be supplemented by two ancilla qubits.}
    \label{tab:H2O_threshold_energies}
\end{table}

We briefly compare resource estimates for our state-preparation protocol with previously reported techniques. If we use a quantum walk to generate a uniform superposition of $D$ bitstrings over $r+1$ qubits, followed by an application of our Walsh circuits, we can prepare the desired wavefunction with exactly $D$ parameters and $\mathcal{O}(D)$ CNOT gates from the Walsh \textit{Ansatz} and $\mathcal{O}(rD)$ CNOT gates from the quantum walk. The most similar approach is known as sum-of-Slaters, which uses $r-1$ ancillary qubits to recursively load the desired initial state. The number of parameters remains $D$, as in the present algorithm; however, the algorithm requires $(r-1)(D-1)$ Toffoli gates.~\cite{Tubman:2018} Alternative approaches for encoding matrix product states (MPSs) have also been reported, most recently with a scaling of $\mathcal{O}(\sqrt{M})$ Toffoli gates, with $\mathcal{O}(\sqrt{2^{r-1}})$ ancillary qubits, where $M$ is the bond dimension of an MPS.~\cite{Berry:2025} Finally, one can direcly prepare the desired state with the quantum walk; however, we found in testing that the gate complexity was considerably worse than the complexity using the Walsh approach. We expect that quantum walks preparing uniform superpositions or states with symmetry are considerably more efficient than those preparing arbitrary states of the same dimension.~\cite{Gonzales:2025} \textcolor{black}{In general, our algorithm is asymptotically comparable with sparse-state preparation, which has a CNOT scaling of $\mathcal{O}(rD)$ and one-qubit-gate scaling of $\mathcal{O}(D\log(D+r))$.~\cite{Greene-Diniz:2026, Gleinig:2021, Gonzales:2025}}

\textit{Conclusions and Discussion--} Here, we present a simple mapping of CI wavefunctions to quantum circuits, which is extended to a variational quantum algorithm for the ground-state energies of molecular Hamiltonians. This algorithm has two key steps, the first is preparing a state of physically-relevant bitstrings and the second is transforming the amplitudes of each SD with the truncated set of Walsh operators. After selecting the $D$ desired CI coefficients, we increase the parameter space to generate a full-rank transformation. A full-rank transformation can be generated by including $D\textrm{log}D$ randomly selected Walsh functions for the partial WFT.~\cite{Tropp:2011} The wavefunction \textit{Ansatz} can be prepared with $\mathcal{O}(D\textrm{log}D)$ CNOT gates, and an $\mathcal{O}(D^2\textrm{log}^2D)$ classical operation. Furthermore, the \textit{Ansatz} does not form a 2-design because it is diagonal and commutative, which mitigates the emergence of barren plateaus,~\cite{McClean:2016, Cerezo:2021, Ragone:2024, Qi:2026} \textcolor{black}{though quantum variational optimization is generally susceptible to flat regions of the potential surface, heavily dependent on the \emph{Ansatz}.~\cite{Cerezo:2021} A more complete understanding of the convergence behavior of VQE and other quantum algorithms with the Walsh \emph{Ansatz} is currently under investigation.}

The subspace preparation step of the algorithm is very flexible. The only requirement is that a uniform superposition is created between the bitstrings of the subspace of interest, while all other amplitudes are zero. Through subspace selection, one can impose other symmetries such as parity or translational symmetry in condensed matter models.~\cite{Altland:1997, Arovas:2022} Effective application of symmetries remains an important tool in classical many-body methods, and the subspace selection protocol can act to symmetry-adapt wavefunctions in quantum algorithms.~\cite{Gard:2020, Lacroix:2020, Selvarajan:2022, Nepomechie:2024}

We proposed two ways to map CI coefficients to Walsh coefficients. We can preserve surjectivity through either oversampling the number Walsh function or using a QR decomposition to find a full rank transformation, which has a maximum classical cost scaling $\mathcal{O}(D^3)$.~\cite{Rupp:2006, Christofides:2014, Bard:2006eprint, Camarero:2018} \textcolor{black}{While diagonalization of the smallest eigenvector of the Hamiltonian can be done in essentially $\mathcal{O}(D)$ complexity, the bottleneck of a given select CI algorithm is dependent on the design details.$^{85-92}$ In quantum approaches the Hamiltonian is written in qubit form, essentially trading matrix manipulations for quantum measurements.}~\cite{Kanno:2023, Robledo-Moreno:2025, Barison:2025, Yu:2025, Sugisaki:2025, Danilov:2025, Sharma:2017aa, Tubman:2020aa} Quantum SCI diagonalizes a subspace CI Hamiltonian but the quantum portion is not iterative, while our algorithm, like all VQEs, is sensitive to the details of the iterative optimization.~\cite{Fedorov:2022, Reinholdt:2025, Tilly:2022aa} For large-scale simulations, we can avoid the QR decomposition by oversampling the Hilbert space by $\mathcal{O}(D\textrm{log}D)$, at a log-linear increase in the number of CNOT gates and parameters. Our algorithm allows for the encoding and quantum variational solution of general CI wavefunctions for electronic structure Hamiltonians.

SCI wavefunctions can also be constructed by alternative means other than the simple threshold scheme used here, for example, by heat-bath or stochastic CI.~\cite{Holmes:2016, Zhang:2021, Rissler:2006} Another possibility is preparing a high-quality guess wavefunction with the Walsh \textit{Ansatz}, followed by application of qubit or fermionic cluster operators,~\cite{McClean:2016, Tang:2021, Yordanov:2021, Anastasiou:2024, Ramoa:2025} resulting in quantum formulations of MR coupled cluster and perturbation theories.~\cite{Bartlett:2007, Angeli:2001} 

Quantum and classical simulation of the all-electron correlation remains a challenge due to the complexity of the correlated reference state, and our approach provides an important sub-routine for these algorithms. \textcolor{black}{We demonstrated our approach using MRCISD, and SCI methods in a VQE context; however, our algorithm could also be used as the starting point for other quantum corrections to dynamical correlation. Quantum algorithms for dynamical correlation corrections have received less attention, in spite of their importance for quantitative accuracy.~\cite{Wu:2025}} Taken together, our combined subspace preparation and Walsh function sampling approach provides a scalable and systematically improvable approach to quantum simulation of correlated many-body ground states.

\textit{Acknowledgements--} KHM acknowledges start-up funding from the University of Minnesota. ARS acknowledges a National Science Foundation (NSF) REU summer fellowship (CHE-2349246). All authors acknowledge the computational resources provided by the Minnesota Supercomputing Institute (MSI) at the University of Minnesota. The authors acknowledge the use of IBM Quantum Credits for this work. 

\textit{Code and Data Availability--} The data from all results, along with the Python code to generate the data, are available at the Zenodo data repository.~\cite{aydogan:QCI}

\bibliography{main} 

\end{document}